\documentclass[11pt]{article}
\usepackage{moriond,epsfig}
\usepackage{graphicx}

\def\be{\begin{equation}}
\def\ee{\end{equation}}
\def\bea{\begin{eqnarray}}
\def\eea{\end{eqnarray}}

\newcommand{\Fi}[1]{Fig.~\ref{#1}}

\newcommand{\agev}{\mbox{~$A$GeV}}

\newcommand{\rb}[1]{\mbox{\textrm{\scriptsize #1}}}

\newcommand{\sqrts}{\ensuremath{\sqrt{s_{_{\rb{NN}}}}}}
\newcommand{\ybeam}{\ensuremath{y_{\rb{beam}}}}
\newcommand{\lam}{\ensuremath{\Lambda}}
\newcommand{\lab}{\ensuremath{\bar{\Lambda}}}
\newcommand{\myphi}{\ensuremath{\phi}}
\newcommand{\pimin}{\ensuremath{\pi^-}}
\newcommand{\piplus}{\ensuremath{\pi^+}}
\newcommand{\pipm}{\ensuremath{\pi^{\pm}}}
\newcommand{\kmin}{\ensuremath{\textrm{K}^-}}
\newcommand{\kplus}{\ensuremath{\textrm{K}^+}}
\newcommand{\kpm}{\ensuremath{\textrm{K}^{\pm}}}

\newcommand{\mypt}{\ensuremath{p_{\rb{t}}}}

\newcommand{\mt}{\ensuremath{m_{\rb{t}}}}

\newcommand{\meanmtm}{\ensuremath{\langle m_{\rb{t}} \rangle - m_{\rb{0}}}}

\newcommand{\gams}{\ensuremath{\gamma_{\rb{S}}}}

\begin{document}
\vspace*{4cm}

\title{ RECENT RESULTS FROM NA49 }

\author{ C. BLUME FOR THE NA49 COLLABORATION }

\address{ Institut f\"ur Kernphysik, J.W.~Goethe Universit\"at Frankfurt, \\
Max-von-Laue Str. 1, 60438 Frankfurt am Main, Germany } 

\maketitle\abstracts{New results of the NA49 collaboration on strange 
particle production are presented. Rapidity and transverse mass 
spectra as well as total multiplicities are discussed.
The study of their evolution from AGS over SPS to the highest RHIC 
energy reveals a couple of interesting features. These include a 
sudden change in the energy dependence of the \mt-spectra and of the
yields of strange hadrons around 30\agev.
Also, fluctuations of the $(K^{+} + K^{-})/(\pi^{+} + \pi^{-})$ ratio and 
the $(p + \bar{p})/(\pi^{+} + \pi^{-})$ ratio, as well as the $v_{2}$ of
\lam\ in Pb+Pb collsions at 158 \agev\ are discussed.}

\section{Introduction}

In the recent years the NA49 experiment has collected data on Pb+Pb 
collisions at beam energies between 20 to 158\agev\ with the objective 
to cover the critical region of energy densities where the expected 
phase transition to a deconfined phase might occur in the early stage 
of the reactions. 
NA49 is a fixed target experiment at the CERN SPS.
Details on the experimental setup can be found in \cite{na49nim}.

\section{Rapidity and transverse mass spectra}

\begin{figure}[htb]
\begin{center}
\includegraphics[height=55mm]{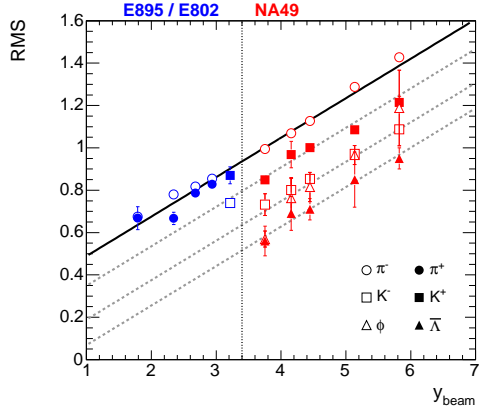}
\end{center}
\caption[]{The RMS values of the rapidity distributions of \pipm, 
\kpm, \myphi, and \lab\ in central Pb+Pb (Au+Au) collisions as a 
function of \ybeam. AGS data are taken from \cite{agspi,agspi2}.
The solid line is a linear fit to the pion data. The dashed
lines have the same slope, but shifted to match the other 
particle species.}
\label{rmsybeam}
\end{figure}

An increase of the RMS-widths of the rapidity spectra with beam energy 
can be observed which for the pions exhibits to a good approximation a linear 
dependence on the beam rapidity in the center-of-mass system \ybeam\ 
over the whole energy range
covered by the AGS and SPS (see \Fi{rmsybeam}). Between 20\agev\
and 158\agev\ this is also true for the other particle types having a
Gaussian-like distribution, with a clear hierarchy in the widths:
$\sigma(\pimin) > \sigma(\kplus) > \sigma(\kmin) \approx \sigma(\myphi) > \sigma(\lab)$.
However, this seems to break down at lower energies, where the
widths of the kaons apparently approach the ones of the pions.

\begin{figure}[htb]
\begin{center}
\includegraphics[width=110mm]{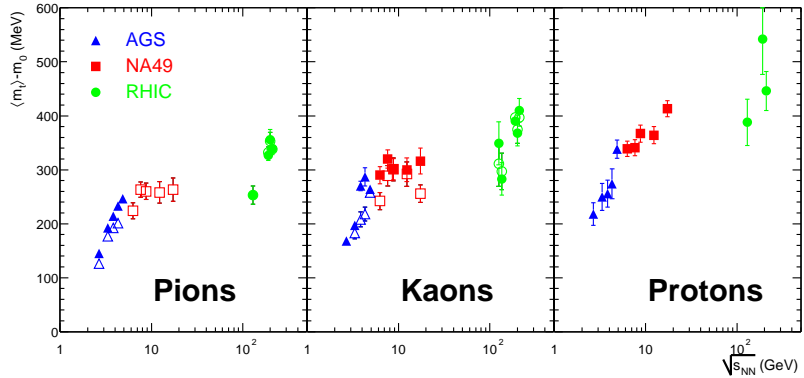}
\end{center}
\caption[]{The energy dependence of \meanmtm\ 
for pions, kaons, and protons at mid-rapidity for 5 (10\%) most central
Pb+Pb/Au+Au reactions. Open symbols represent negatively charged particles.
%AGS and RHIC data are taken from \cite{agspi,agsk,agsp}.
}
\label{meanmt}
\end{figure}

The increase with energy of the inverse slope parameter $T$ of the kaon
\mt-spectra, as derived from an exponential fit, exhibits a sharp
change to a plateau around 30\agev\ \cite{marekqm}. Since the kaon \mt-spectra 
-- in contrast to the ones of the lighter pions or the heavier protons 
-- have to a good approximation an exponential shape, the inverse slope 
parameter provides in this case a good characterization of the spectra.
For other particle species, however, the local slope of the spectra depends
on \mt. Instead, the first moment of the \mt-spectra can be used
to study their energy dependence.
The dependence of \meanmtm\ on the center of mass energy \sqrts\ 
is summarized in \Fi{meanmt}. 
The change of the energy dependence around a beam energy of 
20 -- 30\agev\ is clearly visible for pions and kaons.
While \meanmtm\ rises steeply in 
the AGS energy range, the rise is much weaker from the low SPS energies 
on. To a lesser extent this change is also seen for protons.

\section{Particle yields}

\begin{figure}[t]
\begin{center}
\begin{minipage}[b]{70mm}
\begin{center}
\includegraphics[height=105mm]{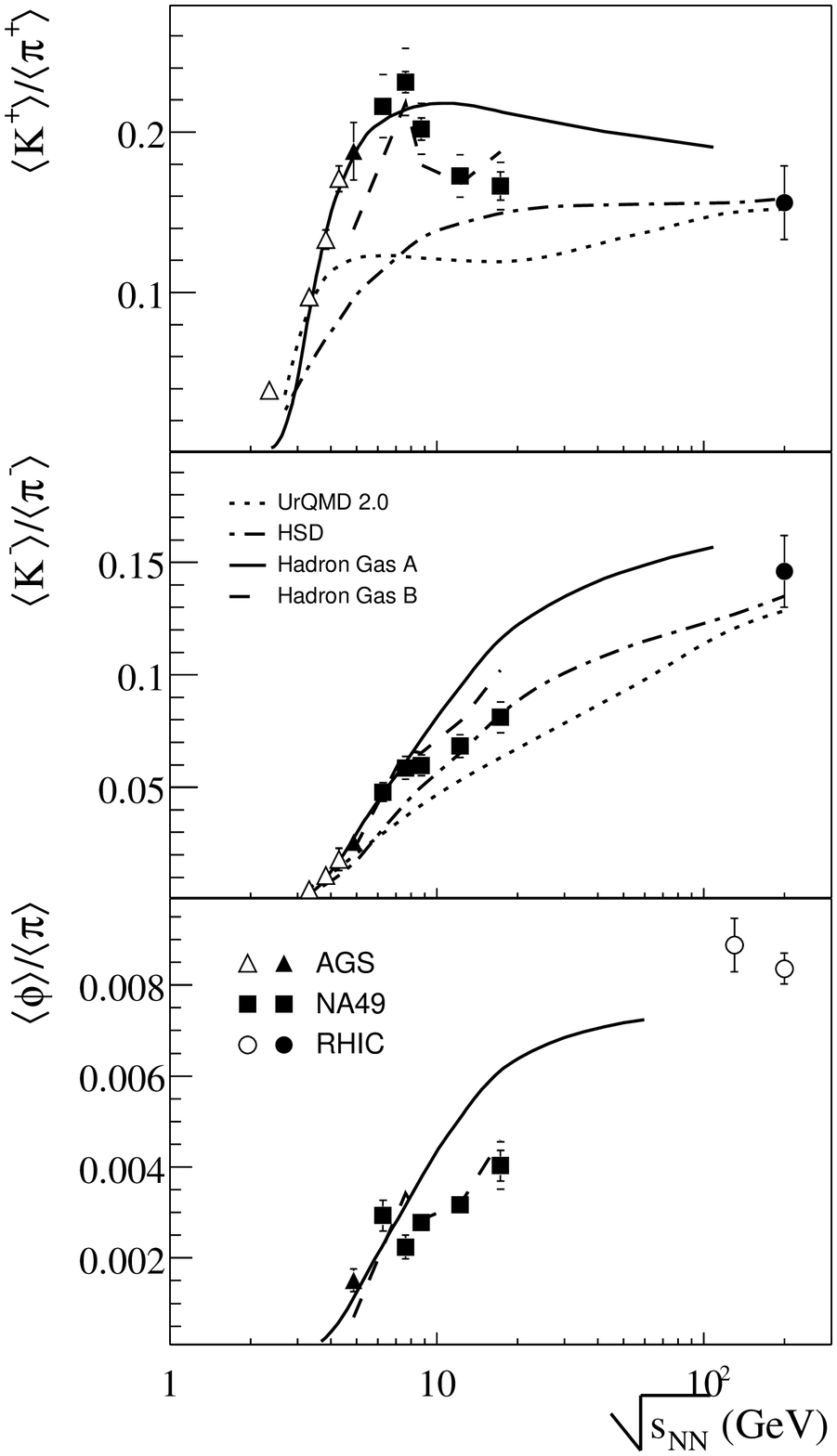}
\end{center}
\end{minipage}
\begin{minipage}[b]{70mm}
\begin{center}
\includegraphics[height=105mm]{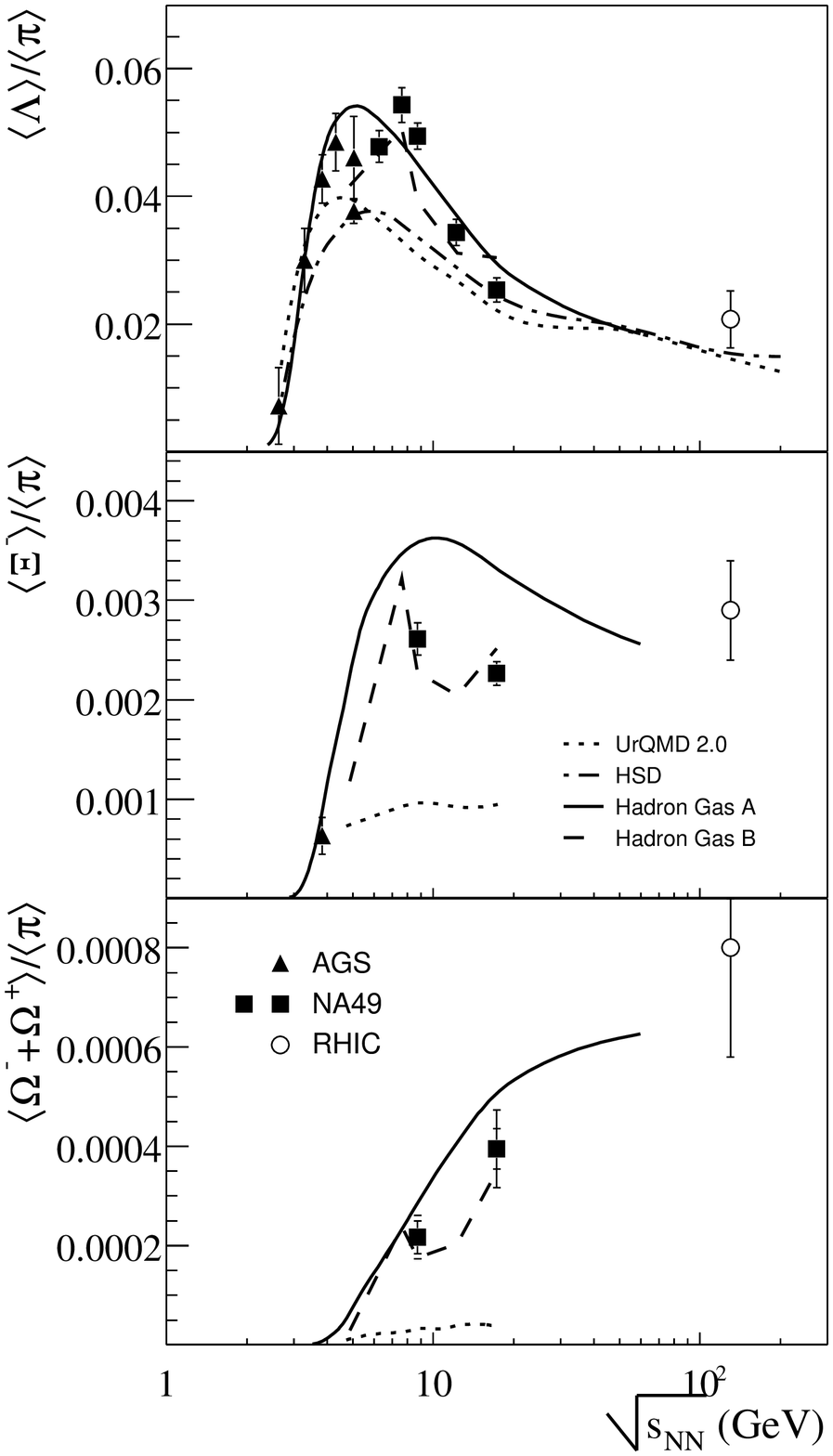}
\end{center}
\end{minipage}
\end{center}
\caption[]{The energy dependence of the 4$\pi$-yields of
strange hadrons, normalized to the pion yields, in central
Pb+Pb/Au+Au collisions. The data are 
compared to string hadronic models \cite{urqmd,hsd} 
(UrQMD 2.0: dotted lines, HSD: dashed-dotted lines) 
and statistical hadron gas models \cite{pbm,becatt}
(with strangeness under-saturation: dashed line, 
assuming full equilibrium: solid line).}
\label{ratios}
\end{figure}

In \Fi{ratios} the energy dependence of the total multiplicities 
for a variety of strange hadrons, normalized to the
pion yield, is summarized and compared to model predictions. 
Generally, it can be stated that the string hadronic models UrQMD and HSD
\cite{urqmd,hsd} do not provide a good description of the data points.
Especially the $\Xi$ and $\Omega$ production is substantially underestimated
and the maximum in the \kplus/\piplus ratio is not reproduced. 
The statistical hadron gas models \cite{pbm,becatt}, on the other hand, 
provide a better overall description of the measurements. However, the 
introduction of an energy dependent strangeness under-saturation factor 
\gams\ is needed \cite{becatt}, in order to capture the structures in the 
energy dependence of most particle species (\kplus, \kmin, \myphi, $\Xi$)
The rapid changed of the hadron production properties observed at low SPS
energies may be related to the onset of deconfinement \cite{marekqm}.

\section{Fluctuations}

\begin{figure}[t]
\begin{center}
\begin{minipage}[b]{50mm}
\begin{center}
\includegraphics[height=55mm]{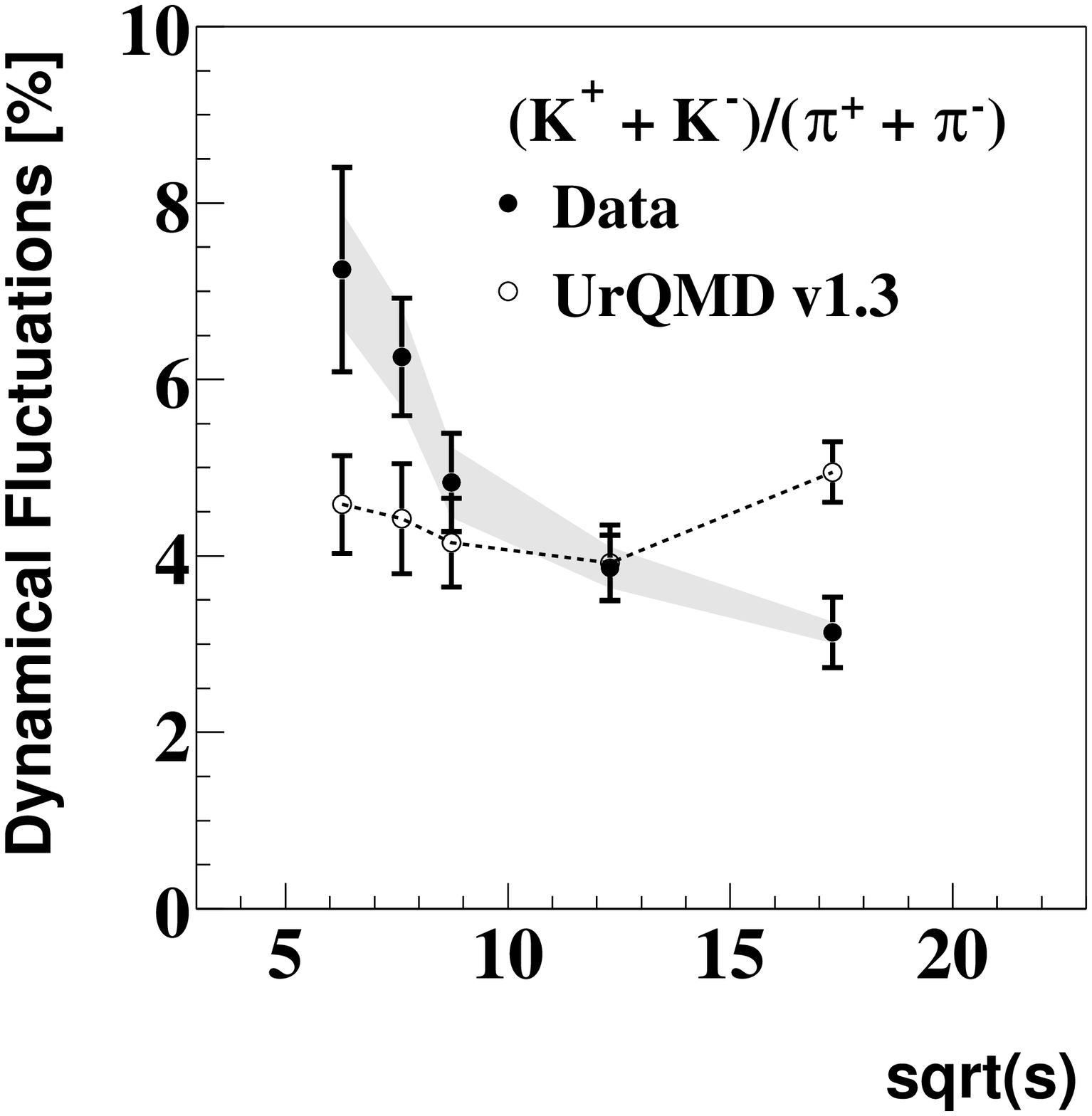}
\end{center}
\end{minipage}
\hspace{10mm}
\begin{minipage}[b]{50mm}
\begin{center}
\includegraphics[height=55mm]{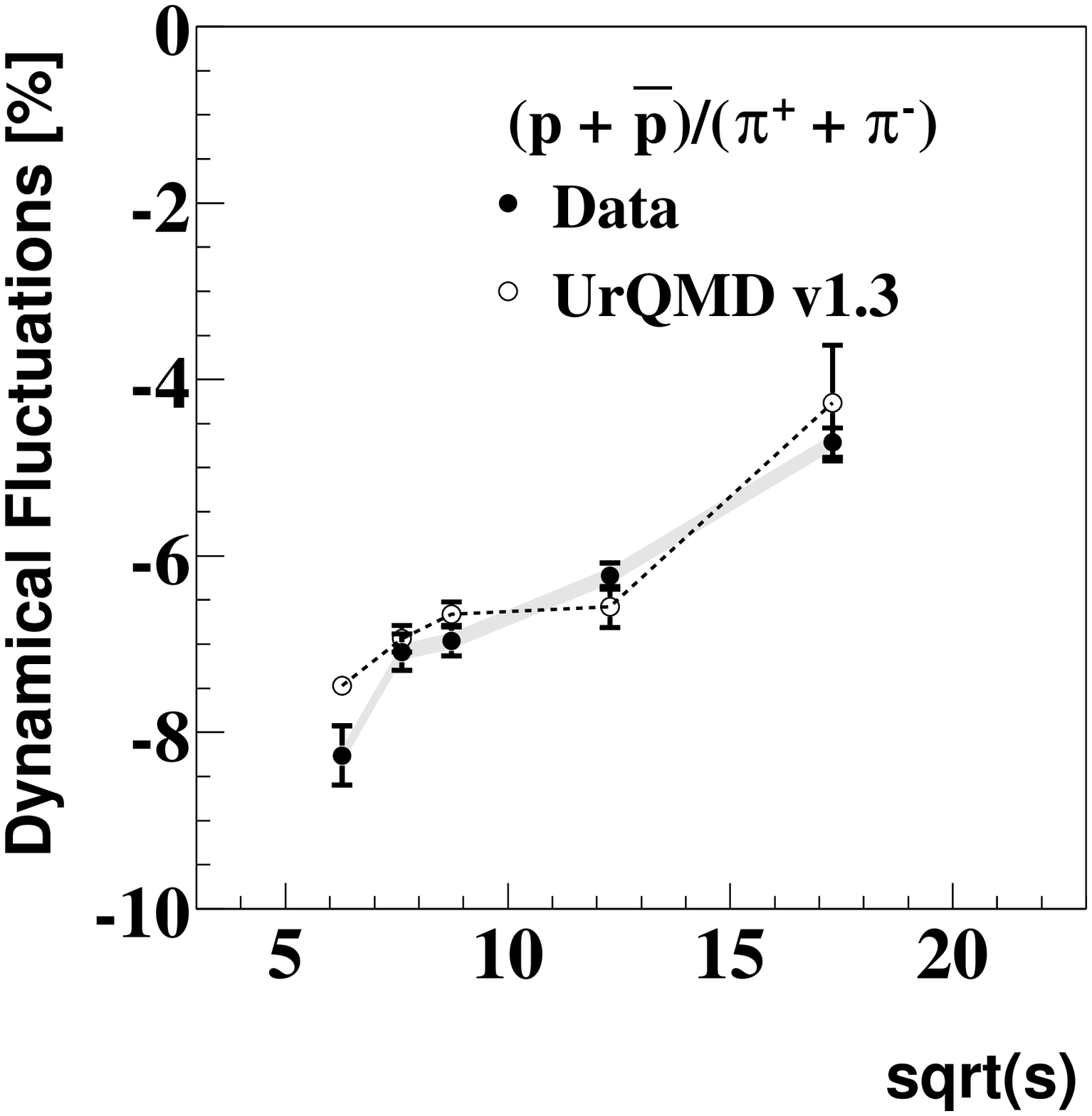}
\end{center}
\end{minipage}
\end{center}
\caption[]{Energy dependence of the event-by-event fluctuation signal of
the $(K^{+} + K^{-})/(\pi^{+} + \pi^{-})$ ratio (left hand side) and
the $(p + \bar{p})/(\pi^{+} + \pi^{-})$ ratio (right hand side) as measured
by NA49 \cite{roland}. The systematic errors of the measurements are 
shown as gray bands.}
\label{kpifluct}
\end{figure}

A study of the energy dependence of event-by-event fluctuations
is given in \Fi{kpifluct}. It shows the fluctuation signal of the 
$(K^{+} + K^{-})/(\pi^{+} + \pi^{-})$ ratio and the 
$(p + \bar{p})/(\pi^{+} + \pi^{-})$ ratio. The dynamical fluctuations
are derived as the difference to a mixed events reference sample
($\sigma_{\rb{dyn.}} = sign(\sigma^{2}_{\rb{data}} - \sigma^{2}_{\rb{mix}}) 
\sqrt( | \sigma^{2}_{\rb{data}} - \sigma^{2}_{\rb{mix}} | ) $).
As shown in the left panel of \Fi{kpifluct}, the $K/\pi$ fluctuations are 
positive and decrease with beam energy. The $p/\pi$ fluctuations, on the 
other hand, are negative
-- indicating a correlation present in the real data -- and increase with 
beam energies. While the trend of the $K/\pi$ fluctuations is not 
reproduced by UrQMD \cite{urqmd},
it provides a good description of the energy dependence of the $p/\pi$ 
fluctuations. This might indicate that the negative value of the 
fluctuations in this ratio is due to resonance decays.

\section{\lam-Flow}

\begin{figure}[t]
\begin{center}
\begin{minipage}[b]{50mm}
\begin{center}
\includegraphics[height=55mm]{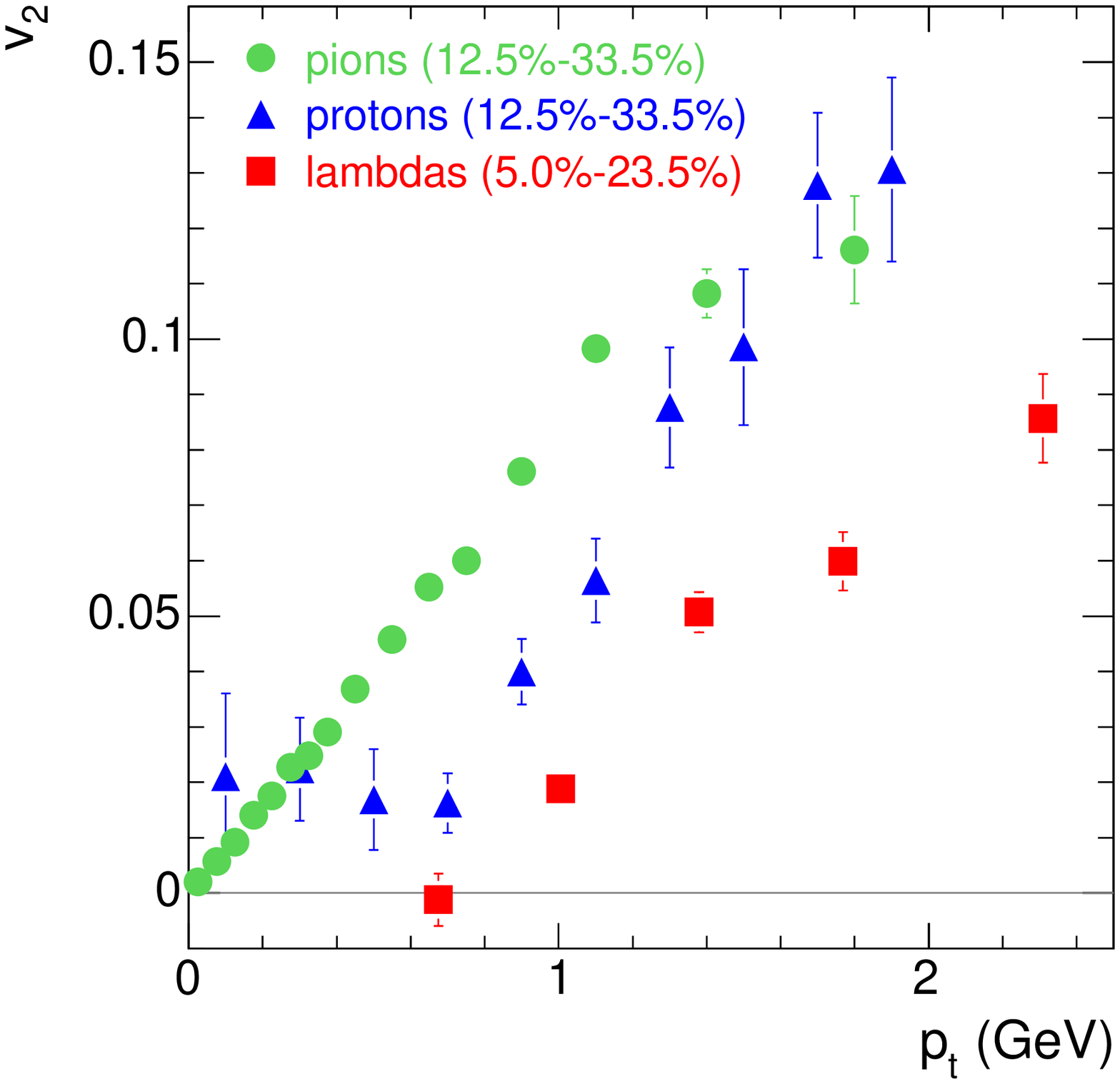}
\end{center}
\end{minipage}
\hspace{10mm}
\begin{minipage}[b]{50mm}
\begin{center}
\includegraphics[height=55mm]{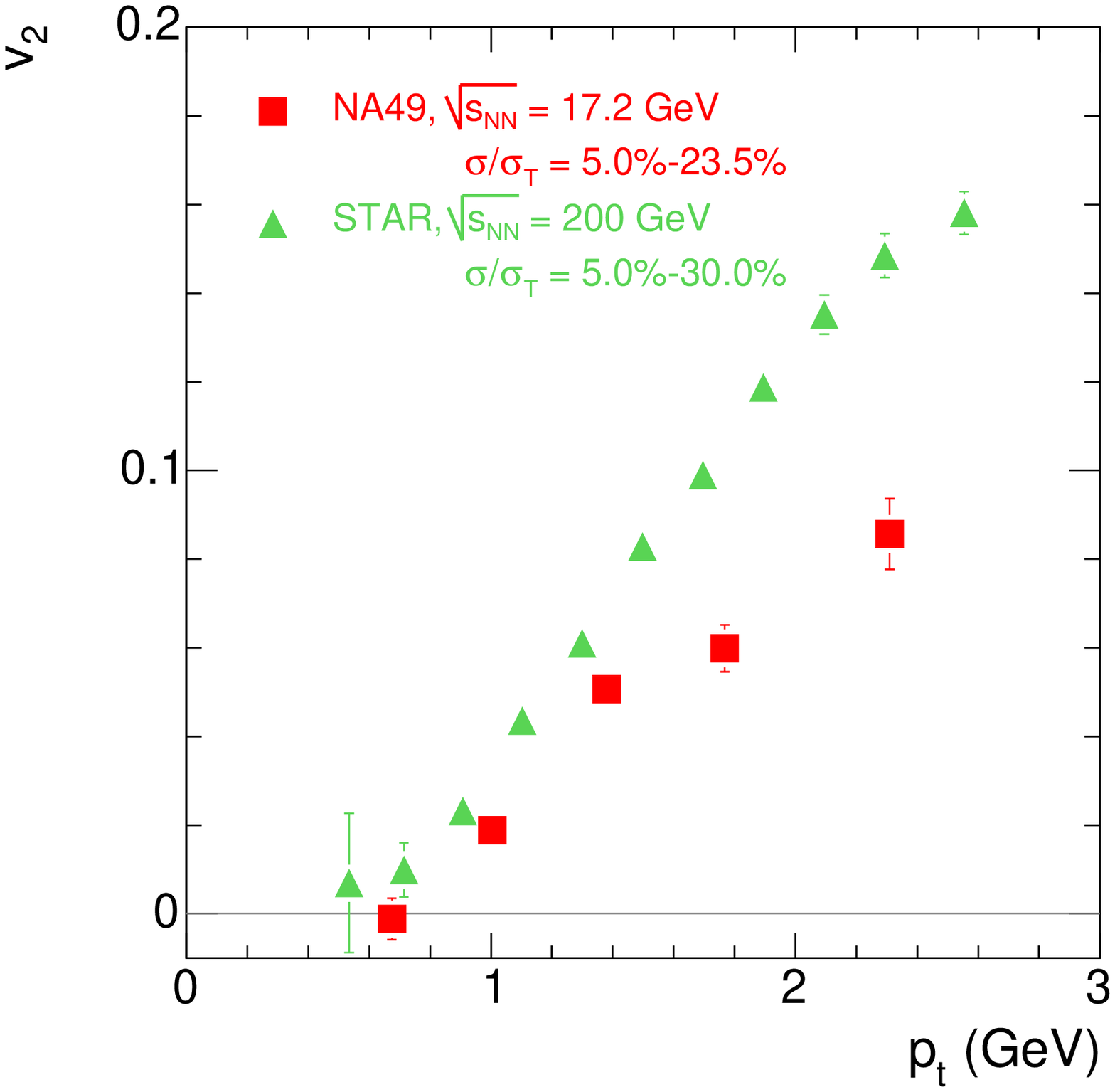}
\end{center}
\end{minipage}
\end{center}
\caption[]{Left: The $v_{2}$ of charged pions, protons, and \lam\ as a function of \mypt\
in semi-central
Pb+Pb collisions at \sqrts~=~17.2 GeV. Please note the difference in centrality selection.
Right: The $v_{2}$ for \lam\ measured at \sqrts~=~17.2 GeV and \sqrts~=~200 GeV \cite{starflow}.}
\label{lamflow}
\end{figure}

Figure~\ref{lamflow} shows the first results on elliptic flow of \lam\ from NA49. The 
comparison to the proton $v_{2}$ reveals a clear difference in the \mypt-dependence.
Also, the increase of $v_{2}$ with \mypt\ is much more pronounced at higher \sqrts\
(see right hand side of Fig.~\ref{lamflow}).

%\section*{Acknowledgments}
%
%This work was supported by the US Department of Energy
%Grant DE-FG03- \\
%97ER41020/A000,
%the Bundesministerium fur Bildung und Forschung, Germany, 
%the Polish State Committee for Scientific Research 
%(2 P03B 130 23, SPB/CERN/P-03/Dz 446/2002-2004, 2 P03B 04123), 
%the Hungarian Scientific Research Foundation (T032648, T032293, T043514),
%the Hungarian National Science Foundation, OTKA, (F034707),
%the Polish-German Foundation, and the Korea Research Foundation 
%Grant (KRF-2003-070-C00015).


\section*{References}
\begin{thebibliography}{99}

\bibitem{na49nim}  S.V.~Afanasiev et al. (NA49 collaboration), 
                   Nucl. Instrum. Meth. {\bf A 430} (1999), 210.

\bibitem{agspi}    J.L.~Klay et al. (E895 Collaboration),
                   Phys. Rev. {\bf C 68} (2003), 054905.

\bibitem{agspi2}   L.~Ahle et al. (E802 collaboration),
                   Phys. Rev {\bf C 57} (1998), 466.

\bibitem{marekqm}  M.~Ga\'zdzicki (for the NA49 collaboration), 
                   J. Phys. {\bf G 30} (2004), S701.

\bibitem{urqmd}    M.~Bleicher et al.,
                   J. Phys. {\bf G 25} (1999), 1859.

\bibitem{hsd}      E.L.~Bratkovskaya et al.,
                   Phys. Rev. {\bf C 69} (2004), 054907.

\bibitem{pbm}      P.~Braun-Munziger, J.~Cleymans, H.~Oeschler, and K.~Redlich,
                   Nucl. Phys. {\bf A 697} (2002), 902.

\bibitem{becatt}   F.~Becattini, M.~Ga\'zdzicki, A.~Ker\"anen, J.~Manninen, and R.~Stock,
                   Phys. Rev. {\bf C 69} (2004), 024905.

\bibitem{roland}   C.~Roland (for the NA49 collaboration), 
                   J. Phys. {\bf G 30} (2004), S1381.

\bibitem{starflow} J.~Adams et al. (STAR collaboration),
                   Phys. Rev. Lett. {\bf 92} (2004), 052302.

\end{thebibliography}
\end{document}